\documentclass[a4paper]{jpconf}
\usepackage{graphicx}
\usepackage{xcolor}
\usepackage{braket}
\usepackage{amsmath}

\graphicspath{{figures/}}

\usepackage{subcaption}
\usepackage[labelfont=bf]{caption}

\usepackage{xurl}

\begin{document}
\title{Impact of quantum noise on the training of quantum Generative
Adversarial Networks}

\author{Kerstin Borras$^{1, 2}$, Su Yeon Chang$^{3,4}$, Lena Funcke$^5$, Michele Grossi$^3$, Tobias Hartung$^{6, 7}$, Karl Jansen$^1$, Dirk Kruecker$^1$, Stefan K\"uhn$^6$, Florian Rehm$^{3,2}$, Cenk T\"uys\"uz$^{1,8}$, and Sofia Vallecorsa$^3$}
\address{$^1$ Deutsches Elektronen-Synchrotron DESY, Platanenallee 6, 15738 Zeuthen, Germany}
\address{$^2$ RWTH Aachen University, Templergraben 55, 52062 Aachen, Germany}
\address{$^3$ CERN, 1211 Geneva 23, Switzerland}
\address{$^4$ Institute of Physics, Ecole Polytechnique F\'ed\'erale de Lausanne (EPFL), 1015 Lausanne, Switzerland}
\address{$^5$ Center for Theoretical Physics, Co-Design Center for Quantum Advantage, and NSF AI Institute for Artificial Intelligence and Fundamental Interactions, Massachusetts Institute of Technology, 77 Massachusetts Avenue, Cambridge, MA 02139, USA}
\address{$^6$ Computation-Based Science and Technology Research Center, The Cyprus Institute, 20 Kavafi Street, 2121 Nicosia, Cyprus}
\address{$^7$ Department of Mathematical Sciences, 4 West, University of Bath, Claverton Down, Bath BA2 7AY, UK}
\address{$^8$ Instit\"ut für Physik, Humboldt-Universit\"at zu Berlin, Newtonstr. 15, 12489 Berlin, Germany}
\ead{su.yeon.chang@cern.ch\\ \textnormal{Preprint number:} MIT-CTP/5400}

\begin{abstract}
Current noisy intermediate-scale quantum devices suffer from various sources of intrinsic quantum noise. Overcoming the effects of noise is a major challenge, for which different error mitigation and error correction techniques have been proposed.
In this paper, we conduct a first study of the performance of quantum Generative Adversarial Networks (qGANs) in the presence of different types of quantum noise, focusing on a simplified use case in high-energy physics. 
In particular, we explore the effects of readout and two-qubit gate errors on the qGAN training process. Simulating a noisy quantum device classically with IBM's Qiskit framework, we examine the threshold of error rates up to which a reliable training is possible. In addition, we investigate the importance of various hyperparameters for the training process in the presence of different error rates, and we explore the impact of readout error mitigation on the results.

\end{abstract}

\section{Introduction}
In classical deep learning, an extensive amount of studies have proven that noise plays a crucial role for the training of neural networks. Artificial noise injection is an efficient regularization method to speed up convergence and to improve the stability of the training process~\cite{Bishop1995, Karras2019}. 

Meanwhile, quantum computing, which is a completely new paradigm of computation, is characterized by statistical uncertainty from its probabilistic nature. Furthermore, on current and near-term quantum hardware one encounters the challenge to overcome the noise due to gate errors, readout errors, and interactions with the environment~\cite{Preskill2018}. The presence of this intrinsic quantum noise suggests the possibility of replacing the artificial noise in the context of classical machine learning with the noise of the quantum hardware. 

In this paper, we study the impact of quantum noise on Quantum Machine Learning (QML), more specifically on qGANs, which are the quantum analog of classical GANs. Focusing on a simplified high-energy physics (HEP) use case, we first explore the impact of readout noise on the qGAN model by classically simulating a noisy quantum device with IBM's Qiskit software development
kit~\cite{Qiskit}. In particular, we study the importance of several training hyperparameters at various levels of noise. We also apply different readout mitigation techniques and study their impact on the performance of qGAN training. Finally, we assess the performance of qGAN training in the presence of the full noise model, including two-qubit gate errors. This work provides broad insights into the impact of noise on QML, which is particularly relevant for applying current noisy intermediate-scale quantum (NISQ) devices~\cite{Preskill2018} to QML. 
\section{Quantum Generative Adversarial Networks}
\label{sec:QGAN}

\begin{figure}[h]
    \begin{subfigure}{0.69\textwidth}
    \centering
    \includegraphics[width = \textwidth]{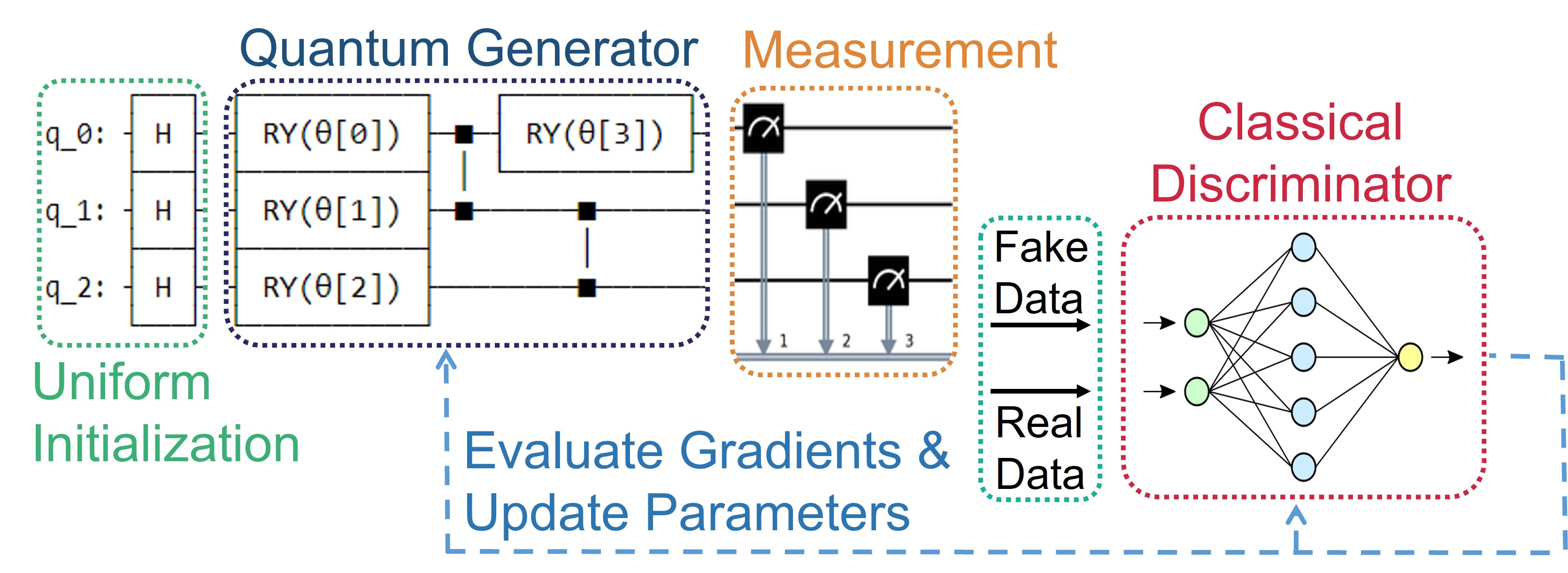}
    \caption{}
    \label{fig:qgan}
    \end{subfigure}
    \hspace{0.015\textwidth}
    \begin{subfigure}{0.29\textwidth}
    \centering
    \includegraphics[width = \textwidth]{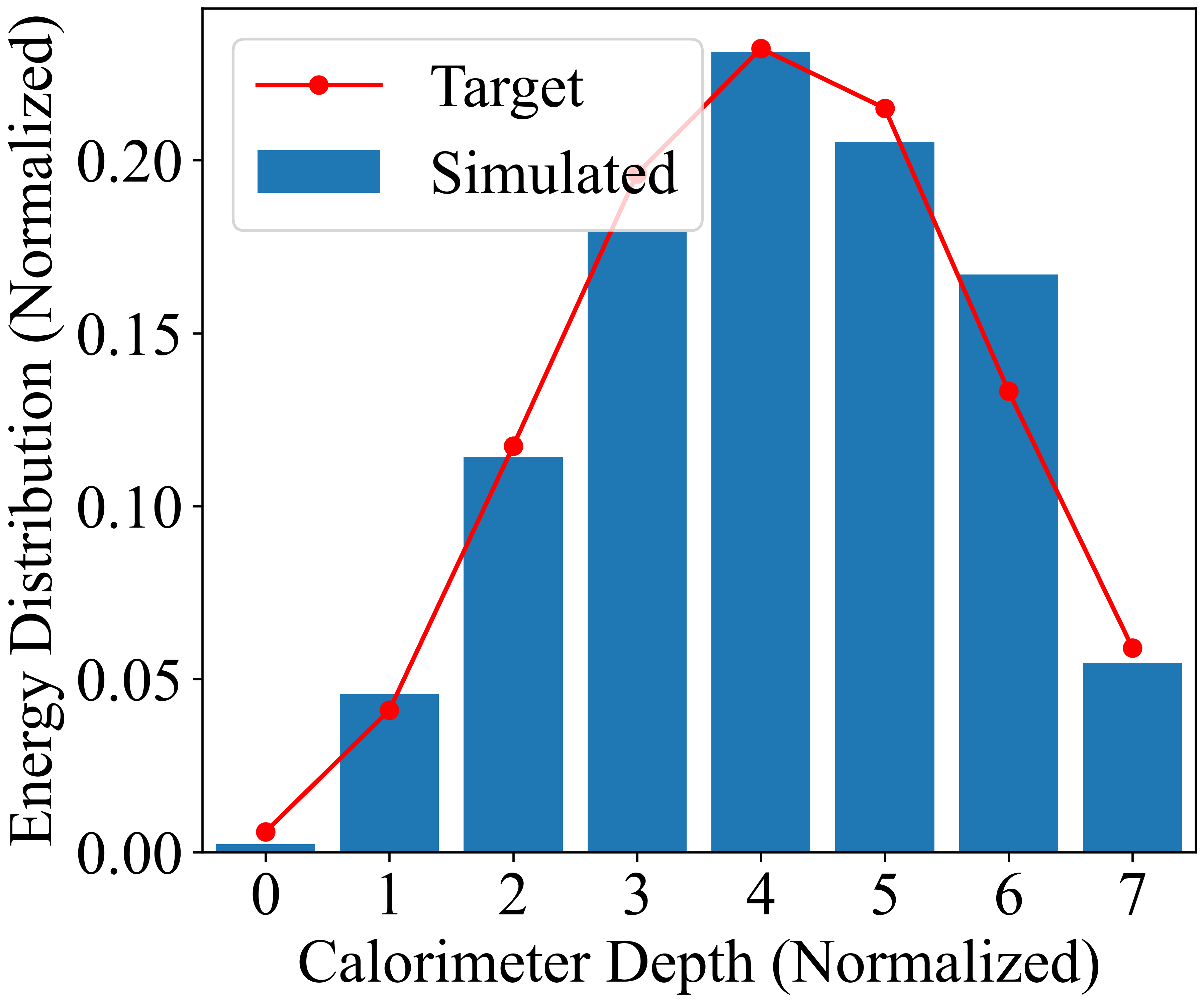}
    \caption{}
    \label{fig:no_noise_simulation}
    \end{subfigure}
\caption{(a) Schematic diagram of the qGAN model, (b) exemplary results of the trained qGAN model applied to a simplified HEP problem using a statevector simulator. 
}
\end{figure}

First, we briefly review the qGAN model introduced in Ref.~\cite{QGAN_qiskit}, which is a hybrid quantum-classical algorithm illustrated in Fig.~\ref{fig:qgan}. This model consists of a classical discriminator and a quantum generator applied to $n$ qubits. The generator is a variational quantum circuit, which consists of alternating layers of single-qubit rotation gates and two-qubit entangling gates. 
In addition, an initial layer of Hadamard gates provides an equal-weight superposition of all computational basis states. By comparing the qGAN performance for different gate choices, we found optimal performance for $R_Y(\theta_i)$ rotation gates and controlled-$Z$ entangling gates. 

The quantum generator is parameterized by the angles $\theta_i$ of the $R_Y(\theta_i)$ gates, while the classical discriminator is parameterized by the neural network parameters $\phi_i$.  During training, the generator and discriminator are optimized alternately. After training, the generator reproduces a probability distribution $Q(x)$ in the computational basis $\ket{x}\in\{\ket{0},..., \ket{2^n - 1}\}$, which approximates the target distribution $P_{\rm target}$. The input for the discriminator contains both real data, which are continuous scalars following the distribution $P_{\rm target}$, and fake data, which are discrete integers between $0$ and $2^n - 1$, chosen by an affine mapping between the computational\linebreak basis states produced by the generator and an equidistant grid over the real data.

In this paper, we use the qGAN model to reproduce electron energy profiles in HEP calorimeters.
To simplify the problem, we sum up the energy distribution along the longitudinal profile, which contains the most essential information to classify the incoming particle. We average over the training set, binning the resulting probability distribution into $2^n = N$ pixels. Figure~\ref{fig:no_noise_simulation} displays the result of an exemplary qGAN training with $n=3$ qubits using a statevector simulator, in the absence of statistical fluctuations and quantum noise.

\section{Quantum Noise Study}
In order to study the impact of quantum noise on the training process, we use a qGAN model with a quantum generator consisting of $n=3$ qubits and $2$ learning layers. Readout errors dominate over two-qubit gate errors for shallow circuits; thus, we choose to focus on the former. We assume that readout errors, i.e., misidentifying a measurement outcome as $1$ given it was $0$ and vice versa, are uncorrelated between different qubits and that the bit-flip probability $p$ for misidentifying the measurement outcome is the same for each qubit. For the following simulations, we evaluate the performance of the qGAN model using the relative entropy
\begin{equation}
D_{\rm KL}(P||Q) = \sum_{x = 0}^{N-1} P(x) \log \frac{P(x)}{Q(x)} ,   
\label{eq:KLdivergece}
\end{equation}
also called the Kullback-Leibler (KL) divergence. Here, $Q(x)$ is the output distribution of the quantum generator and $P(x)$ is the discretized version of the target distribution $P_{\rm target}$.

\subsection{Hyperparmeter Scan}
\label{sec:hyperparameter_scan}
\begin{figure}[h]
\centering
\begin{subfigure}{0.31\textwidth}
    \centering
    \includegraphics[width = \textwidth]{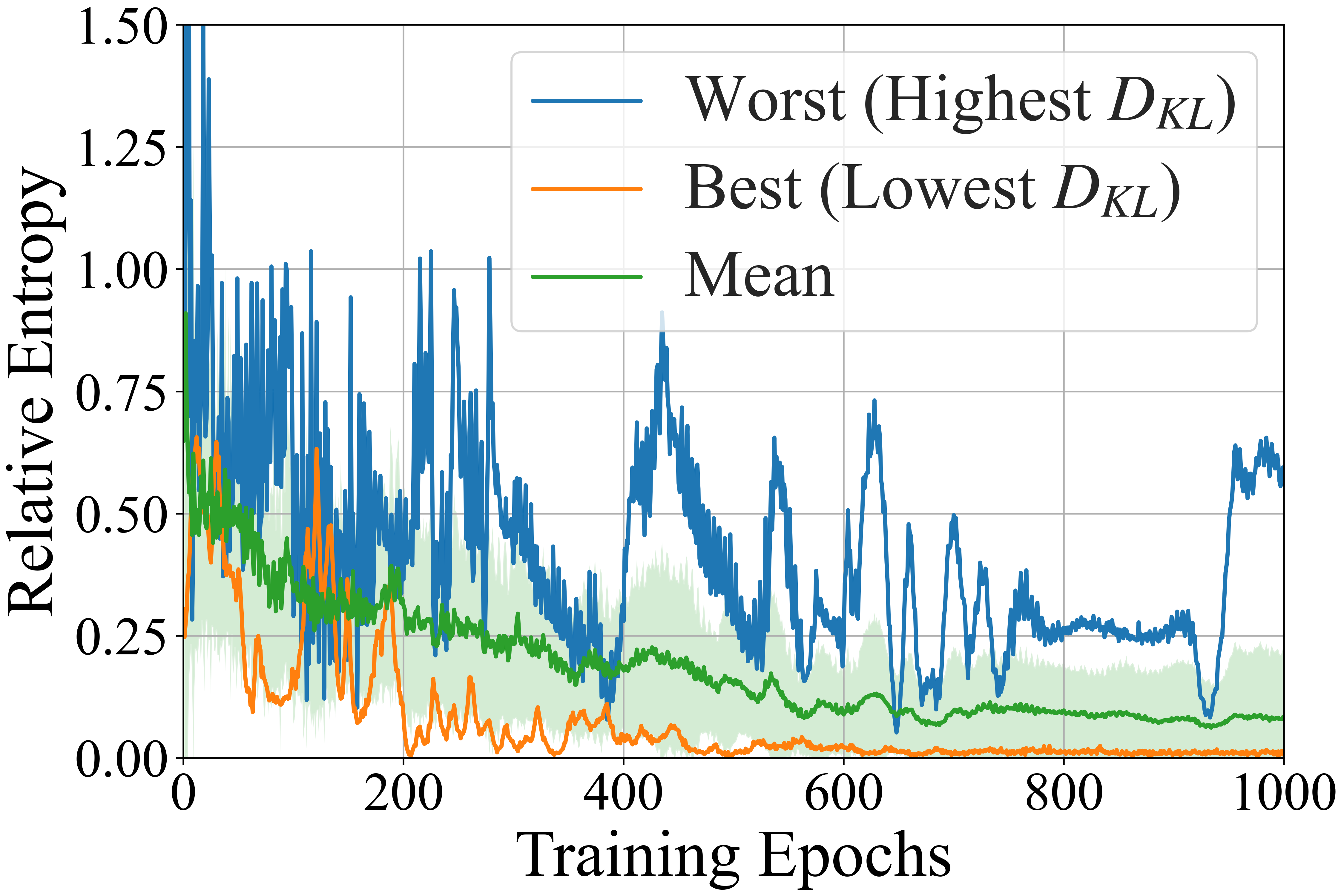}
    \caption{$p=0.01$}
    \label{fig:scan_ro_noise_p001}
\end{subfigure}
\begin{subfigure}{0.31\textwidth}
    \centering
    \includegraphics[width = \textwidth]{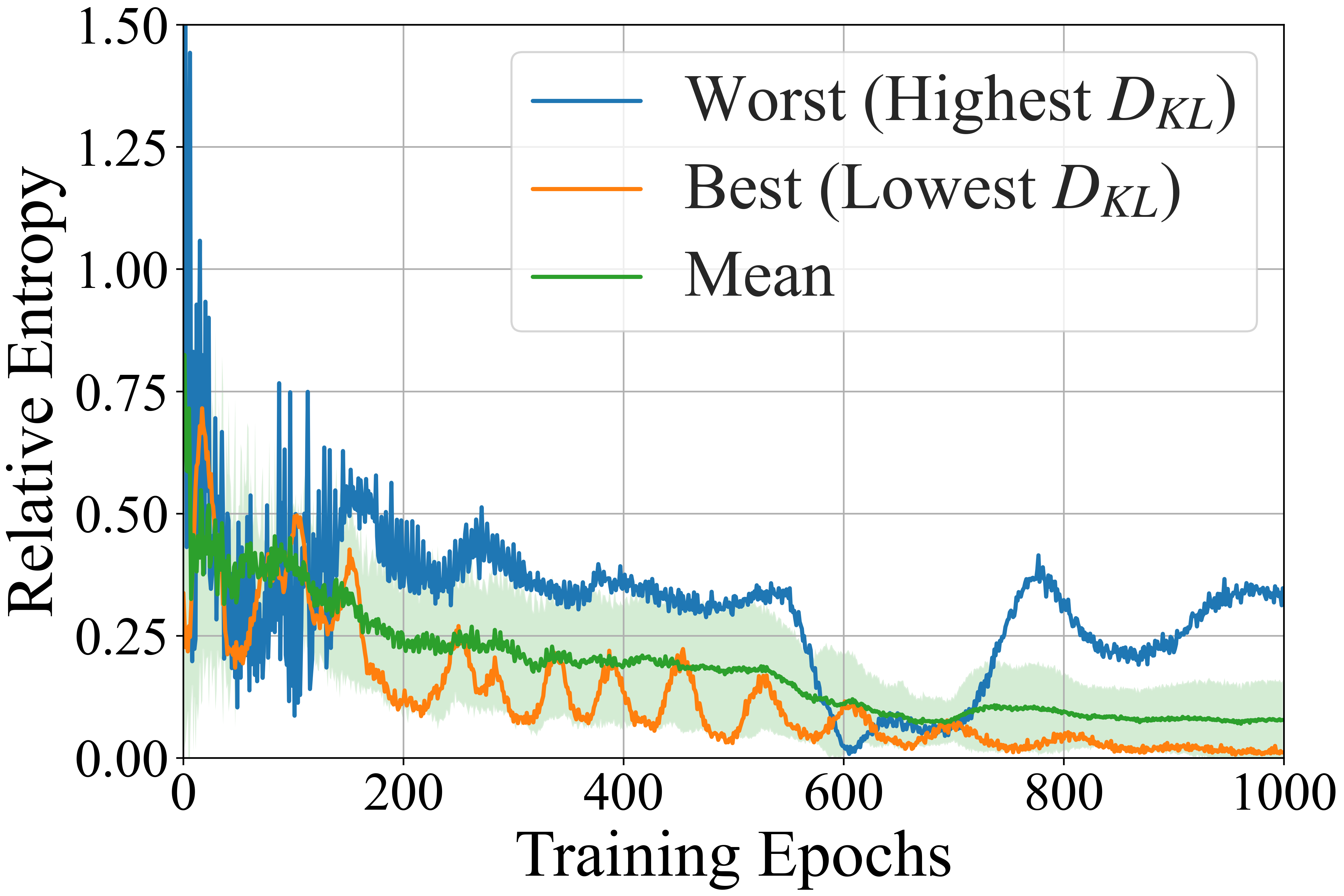}
    \caption{$p=0.05$}
    \label{fig:scan_ro_noise_p005}
\end{subfigure}
\begin{subfigure}{0.31\textwidth}
    \centering
    \includegraphics[width = \textwidth]{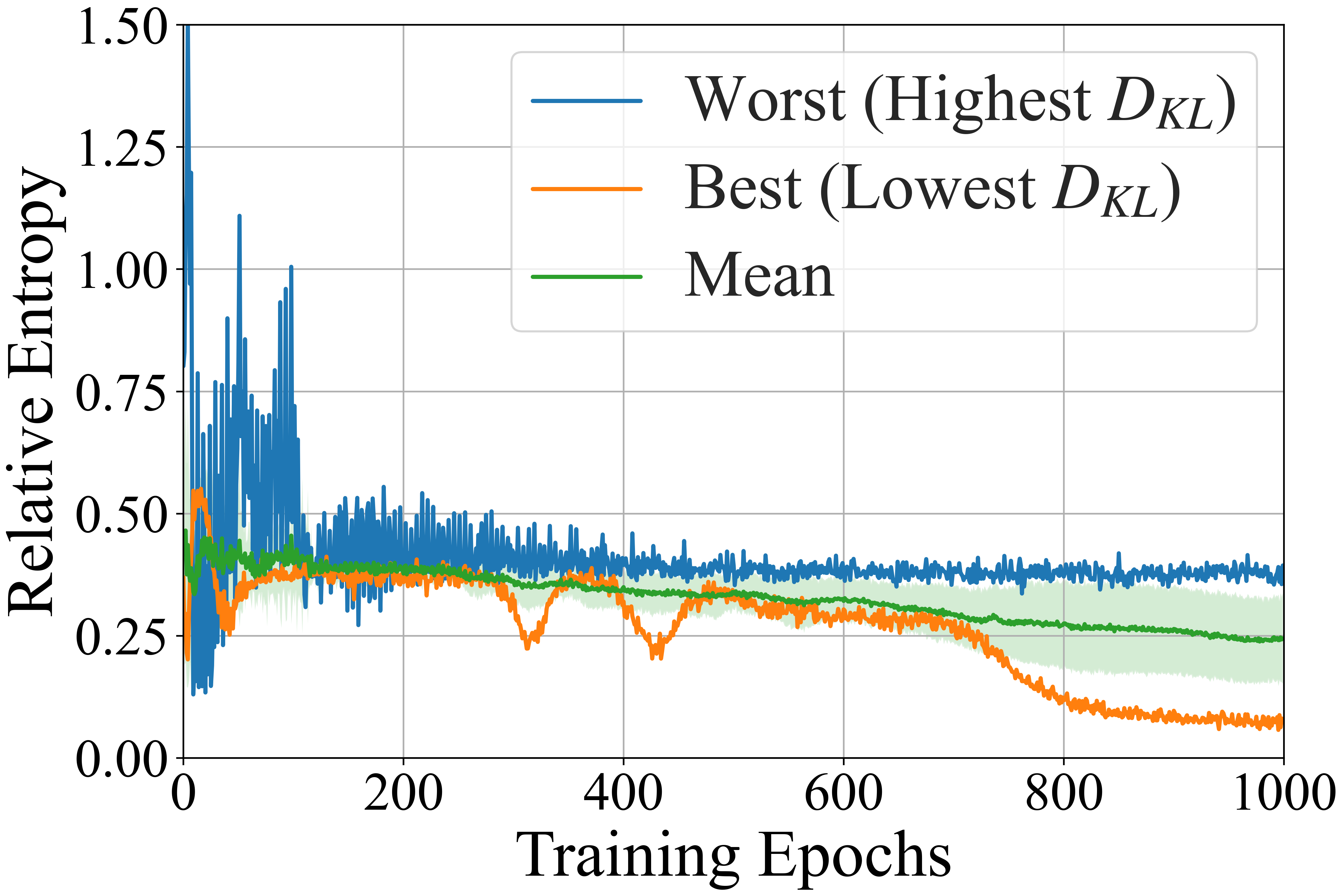}
    \caption{$p=0.1$}
    \label{fig:scan_ro_noise_p01}
\end{subfigure}
\caption{Relative entropy defined in Eq.~\eqref{eq:KLdivergece} as a function of training epochs, shown for different hyperparameter choices and for different bit-flip probabilities $p$. 
}
\label{fig:scan}
\end{figure}

In this section, we study different values of the bit-flip probability, $p=\{0.01,0.05, 0.1\}$, as well as different subsets of hyperparameters: the generator learning rate $lr_g$, the discriminator learning rate $lr_d$, and the the exponential decay rate $\gamma$ for the learning layers. In Fig.~\ref{fig:scan}, we plot our results for the relative entropy using the best (orange) and worst (blue) choices of hyperparameters, as well as an average over all runs (green). For an appropriate choice of hyperparameters, the qGAN model converges for all choices of $p$. However, the relative entropy increases with a higher noise level, which cannot be fully mitigated even when choosing optimal hyperparameters.
\begin{figure}[h]
    \centering
    \includegraphics[width = 0.49\textwidth]{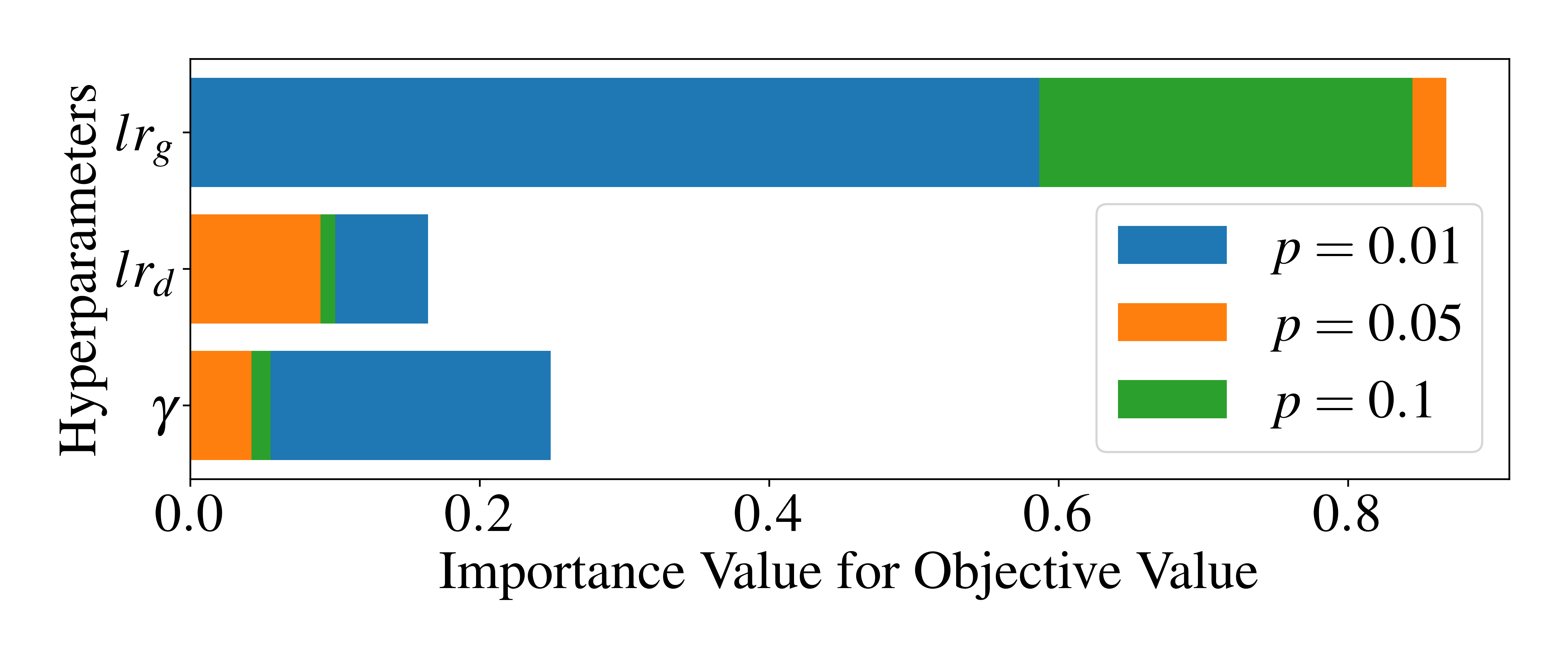}
    \caption{Importance of hyperparameters for different strengths of the readout error.}
    \label{fig:importance}
\end{figure}

To examine which of the three hyperparameters has the strongest impact on the training for different noise levels, we display the so-called importance value of the hyperparameters for the objective value in Fig.~\ref{fig:importance}, which we calculated based on the fANOVA hyperparameter importance evaluation algorithm \cite{favona}.
We find that the generator learning rate $lr_g$ has the highest impact on the training, which demonstrates the difficulty of training the quantum generator. Thus, as expected, the impact of $lr_g$ grows as the bit-flip probability increases.

\subsection{Instability of qGAN Training}
Instabilities, which manifest as large fluctuations of the relative entropy when choosing different initial training parameters, are one of the major challenges that have to be overcome in GAN training, both in the classical and the quantum case. In this section, we investigate the instability of qGAN training in presence of readout noise, using the optimal hyperparameters found in Section~\ref{sec:hyperparameter_scan}. The initial training parameters $\theta_i$ of the quantum generator circuit 
are sampled randomly from a normal distribution with a mean of 0.1 and a standard deviation of 0.2.
\begin{figure}[h]
\centering
    \begin{subfigure}{0.31\textwidth}
        \includegraphics[width = \textwidth]{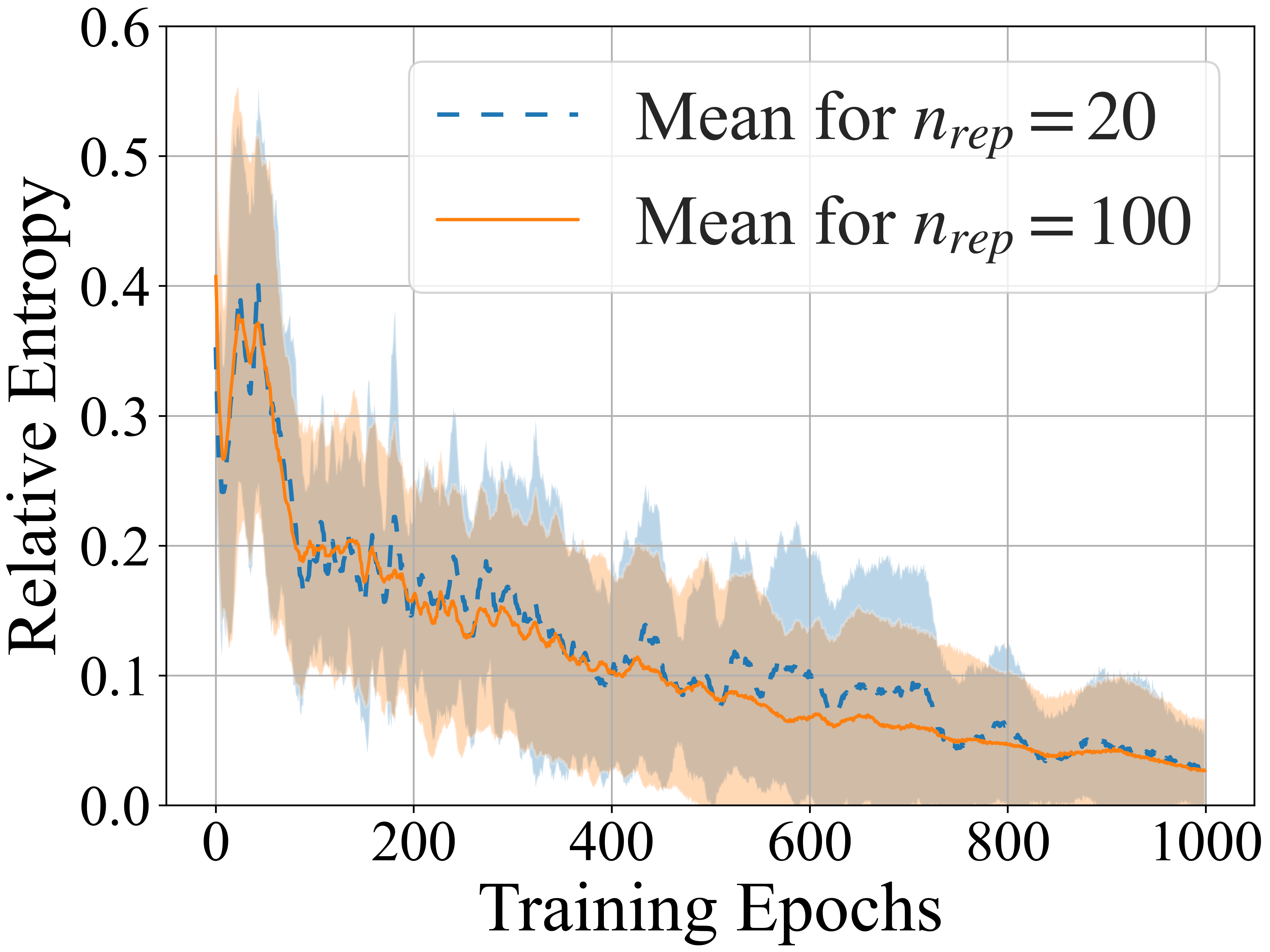}
        \caption{$p = 0.01$}
        \label{fig:1E-2_compare}
    \end{subfigure}
    \begin{subfigure}{0.31\textwidth}
        \includegraphics[width = \textwidth]{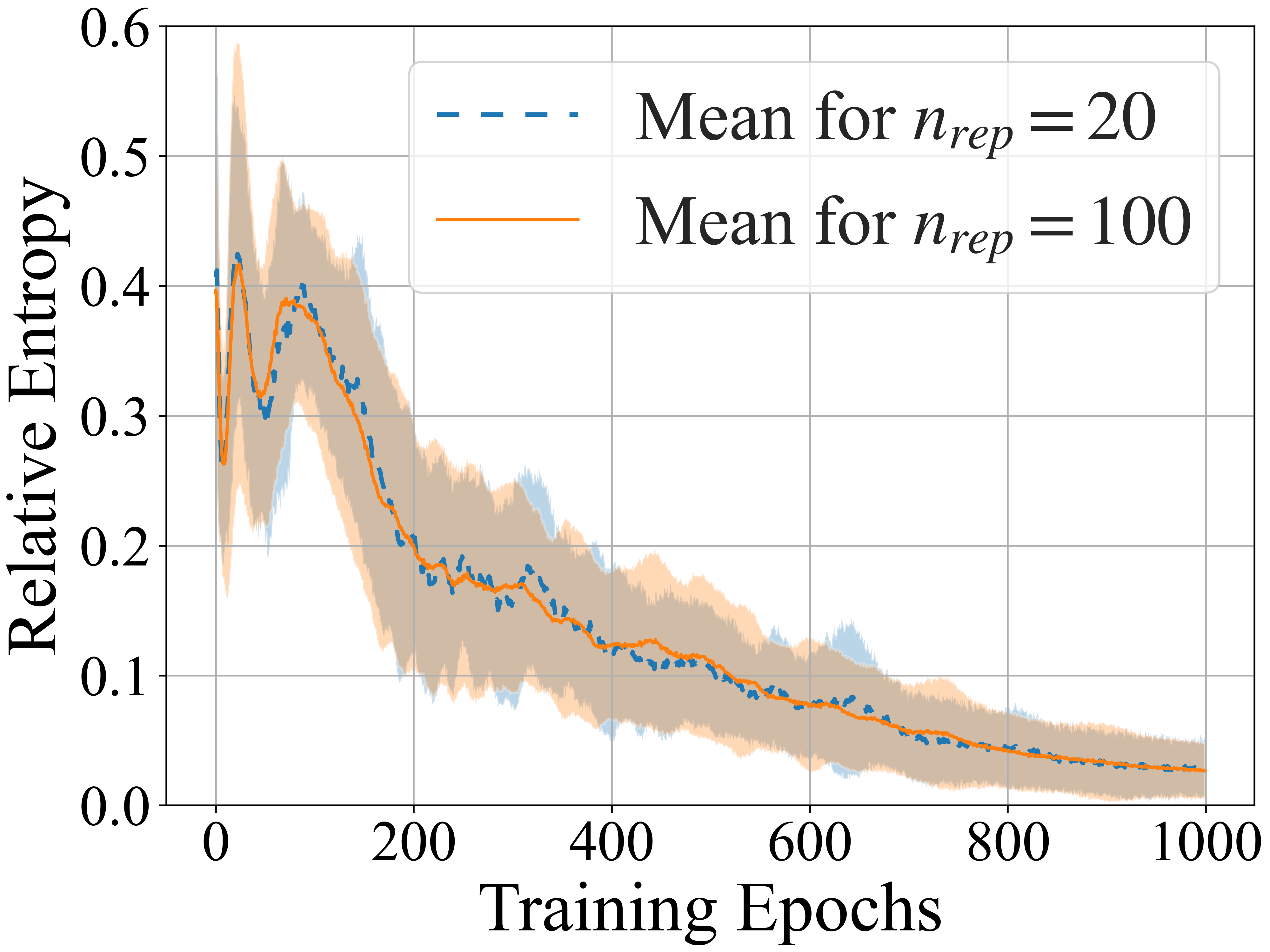}
        \caption{$p = 0.05$}
        \label{fig:5E-2_compare}
    \end{subfigure}
    \begin{subfigure}{0.31\textwidth}
        \includegraphics[width = \textwidth]{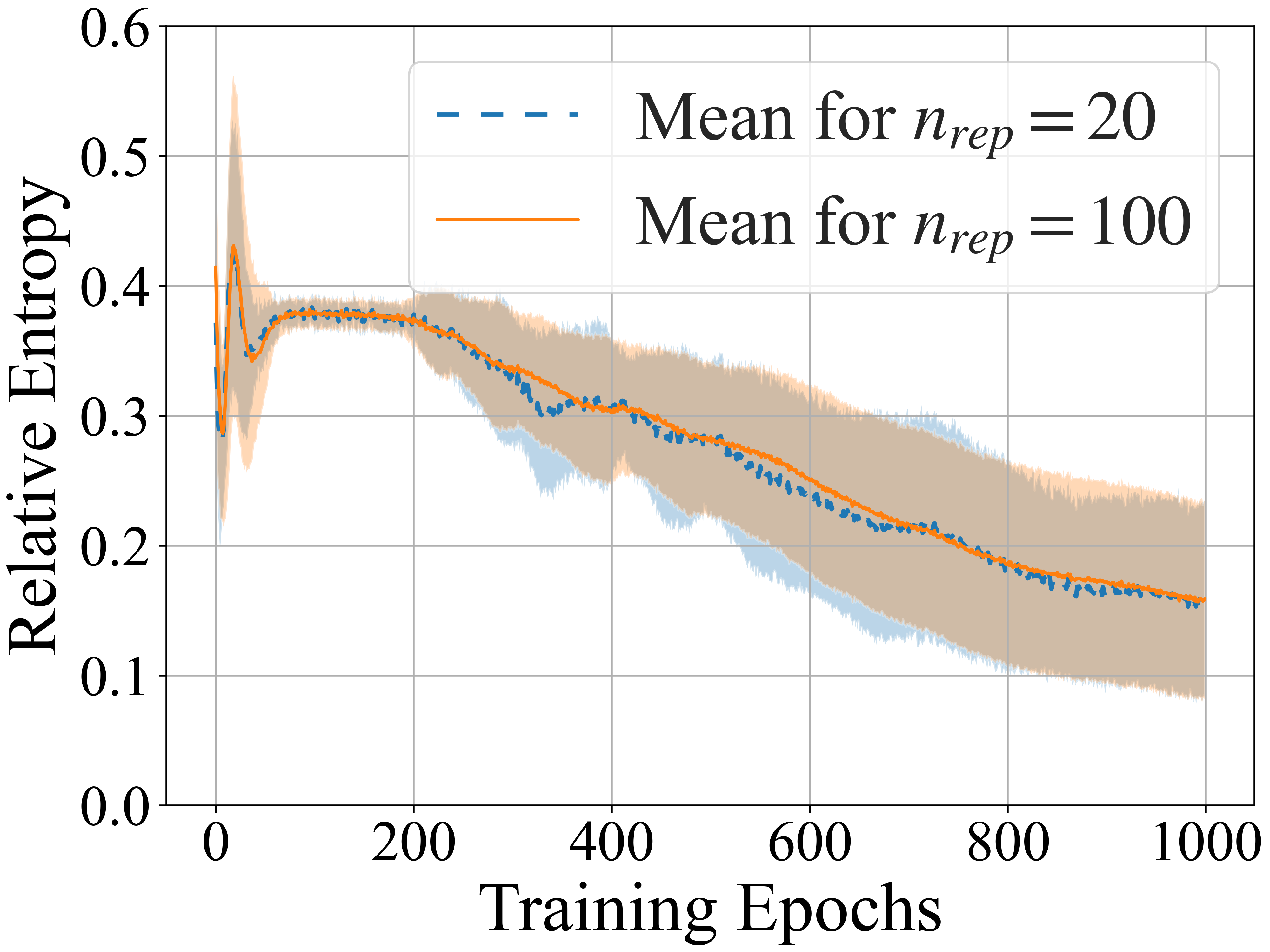}
        \caption{$p = 0.1$}
        \label{fig:1E-1_compare}
    \end{subfigure}
\caption{Relative entropy defined in Eq.~\eqref{eq:KLdivergece} as a function of training epochs, averaged over different numbers of repetitions $n_{\rm rep}$, shown for different bit-flip probabilities $p$. 
}
\label{fig:instability}
\end{figure}

Figure~\ref{fig:instability} summarizes the results of the qGAN training for different numbers of repetitions, $n_{\rm rep}=20$ (blue) and $n_{\rm rep}=100$ (orange), and for different readout-noise levels. 
Interestingly,  for all values of $p$, both the average relative entropy and the standard deviation do not substantially change as we increase the statistics by increasing $n_{\rm rep}$.
 Thus, although the model converges in a stable manner on the ensemble of simulations, individual unstable runs occur with a constant probability. These results demonstrate that training instabilities occur independently from the readout-error level and cannot be alleviated by quantum noise or increasing statistics. 

\subsection{Error Mitigation}

In the following, we study two different methods to mitigate readout errors: conventional bit-flip (BF) mitigation~\cite{qiskit_error_mitigation} and independent bit-flip (IBF) mitigation~\cite{funcke_measurement_2020}. These methods rely on a calibration matrix ($M$), which maps the measured probability distribution to the expected probability distribution ($ P_{\rm noisy} = M P_{\rm ideal}$). In this way, the inverse of $M$ can be used to obtain the expected outcome from measured probabilities. The matrix $M$ for a single qubit reads
\begin{equation}
    M = \begin{bmatrix}
    1-p_{01} & p_{10} \\
    p_{01} & 1-p_{10} 
    \end{bmatrix},
    \label{eq:M}
\end{equation}
where $p_{01}$ ($p_{10}$) is the bit-flip probability of misidentifying 0 as 1 (1 as 0) during readout.

For the conventional BF method~\cite{qiskit_error_mitigation}, the computational resources required to obtain the inverse of $M$ grow exponentially with respect to the number of qubits ($M$ is a $2^n\times 2^n$ matrix). In contrast, the IBF method~\cite{funcke_measurement_2020} assumes that the single-qubit matrix $M$ in Eq.~\eqref{eq:M} can be extended to multiple qubits with a tensor product over all qubits ($M = M_1 \otimes M_2 \otimes \cdots \otimes M_N$). Thus, this method assumes uncorrelated multi-qubit readout errors, which is an assumption that has recently been verified experimentally~\cite{Alexandrou:2021wqu}. This way, computing the inverse of $M$ can be done by computing the inverse of $n$ many $2\times 2$ matrices using polynomial resources.

Correct estimation of the bit-flip probabilities is essential for successful error mitigation. 
Therefore, one needs to execute calibration circuits on the quantum device to obtain these values. The precise estimation of these values might be limited by their time dependence and by the finite sampling size (shot noise), e.g., $\mathcal{O}(1/\epsilon^2)$ shots are required for $\epsilon$ precision. 

\begin{figure}[h]
\centering
\begin{subfigure}{0.42\textwidth}
\centering
    \includegraphics[width = \textwidth]{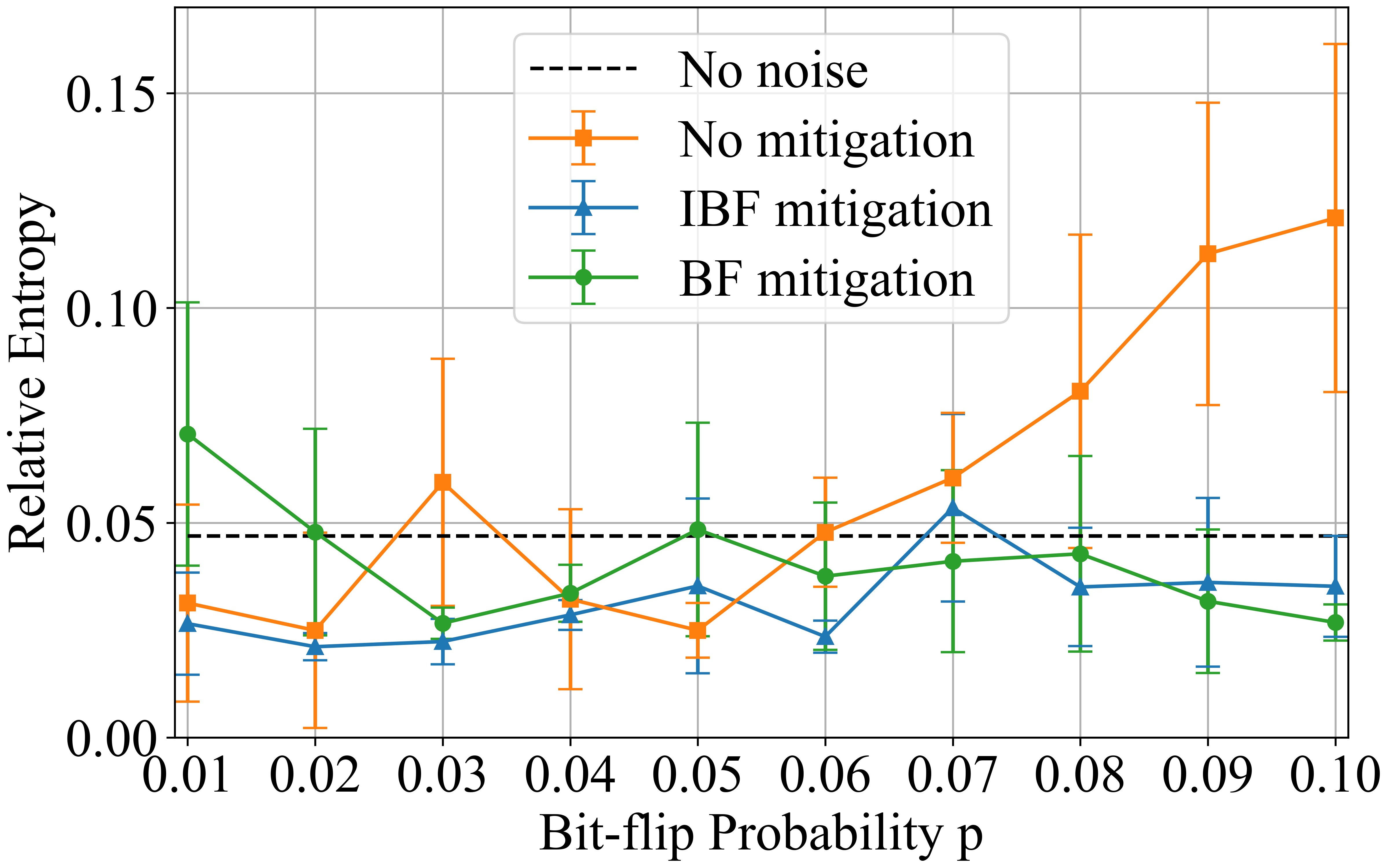}
    \caption{}
    \label{fig:ent_vs_noise}
\end{subfigure}
\begin{subfigure}{0.42\textwidth}
\centering
    \includegraphics[width = \textwidth]{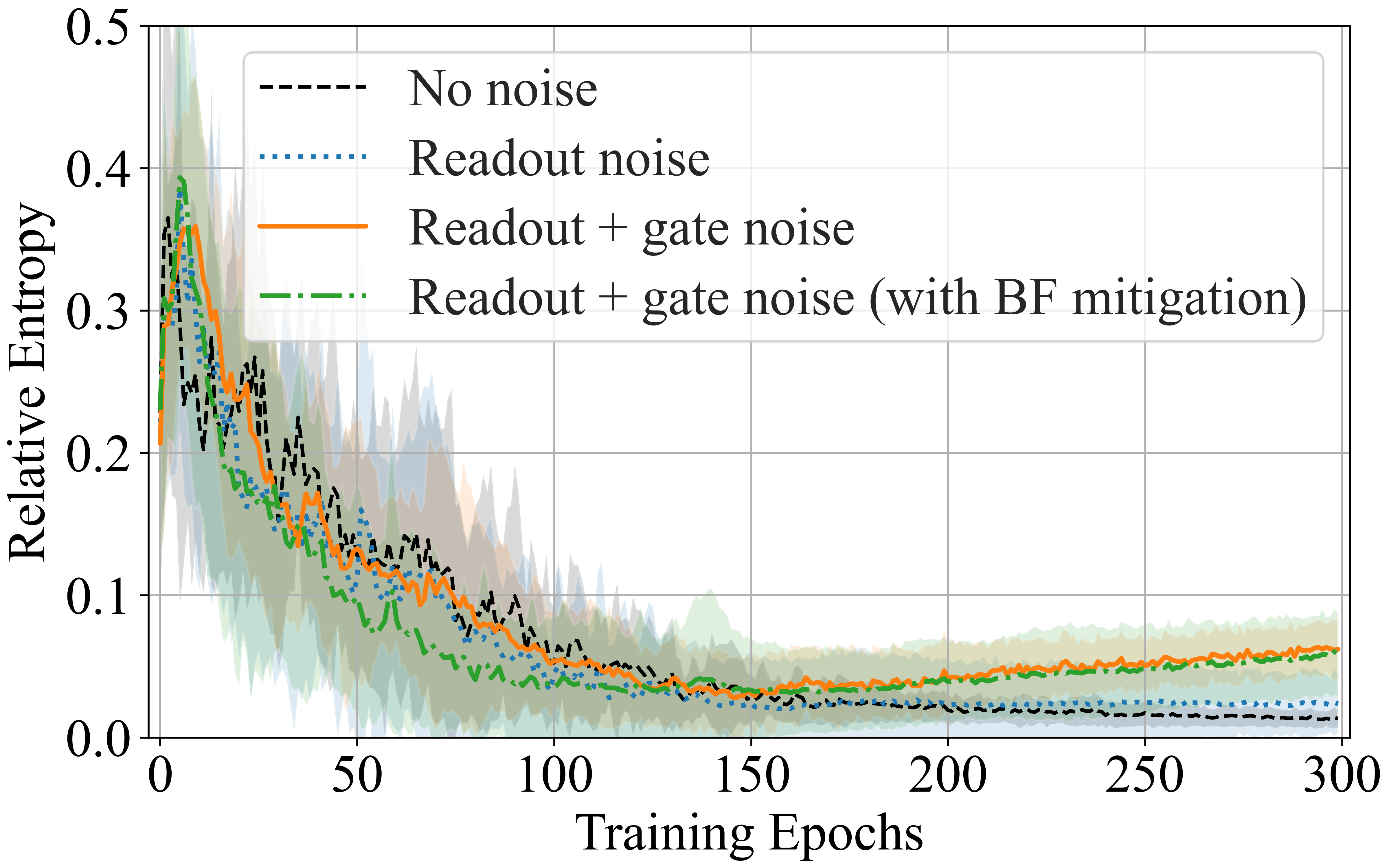}
    \caption{}
    \label{fig:full_noise}
\end{subfigure}  
\caption{(a) Final relative entropy after qGAN training in the presence of readout noise, as a function of the bit-flip probability $p=p_{01}=p_
{10}$, averaged over $n_{\rm rep}=20$ repetitions, shown with and without error mitigation. (b)~Relative entropy as a function of training epochs, shown with different types of noise, with and without readout-error mitigation.}
\label{fig:error_mitigation}
\end{figure}

In Fig.~\ref{fig:ent_vs_noise}, we examine the impact of readout errors and error mitigation on the final relative entropy after qGAN training, comparing the cases of no noise (black), readout noise (orange), IBF mitigation (blue), and BF mitigation (green).
When focusing on the average value of the relative entropy, the results for training with a small readout error ($p < 0.06$) seem to outperform the results without noise, which aligns with our initial suggestion of enhancing the qGAN training with quantum noise. However, within the statistical uncertainty, the noisy results for $p < 0.06$ are compatible with both the noise-free and the error-mitigated results.  For larger levels of readout error ($p \geq 0.06$), we observe that the noisy results become worse compared to the noise-free and error-mitigated cases. This implies that error mitigation plays a crucial role for qGAN training in the presence of the large readout errors on current NISQ devices. 

When comparing the two different error mitigation methods in Fig.~\ref{fig:ent_vs_noise}, we find that the resulting values for the relative entropy agree within the statistical uncertainty, as expected. For small values of $p$, we find a small deviation between the mean values for the relative entropy, which is most likely caused by using exact probabilities to estimate the calibration matrix $M$ in the case of the IBF method and using a finite number of samples (3000 shots) in the case of the BF method. In the latter case, the finite number of samples results in a precision error of $\epsilon\sim \mathcal{O}(0.02)$ for estimating $p=p_{01}=p_{10}$, in agreement with our results in Fig.~\ref{fig:ent_vs_noise}.

Finally, we study the impact of both readout and two-qubit gate errors on the qGAN training. In Fig.~\ref{fig:full_noise}, we compare the results for the relative entropy (orange) to the results for the no-noise case (black), the readout-error-only case (blue), and the case of BF-mitigated readout errors and unmitigated gate errors (green). For this comparison, we consider error rates of 2.5\% for the readout error and 1.5\% for the gate error. As shown in Fig.~\ref{fig:full_noise}, new optimized hyperparameters are found to reduce the number of training epochs to only 300 epochs until reaching convergence. Furthermore, a new training schema is adapted, by updating the quantum generator once and the classical discriminator 10 times at each step. Similar to our findings in Fig.~\ref{fig:ent_vs_noise}, the inclusion of quantum noise does not improve the results for the relative entropy, as the results for the noise-free, noisy, and error-mitigated cases agree within the statistical uncertainty. Interestingly, the average values of the relative entropy quickly converge in the noise-free and readout-error-only cases, but do not seem to converge when including two-qubit gate errors. These findings reveal the critical effect of two-qubit gate errors on qGAN training with current NISQ devices. 

\section{Conclusion}
In this paper, we studied the impact of quantum noise on qGAN training, focusing on a simplified use case in high-energy physics. We first investigated different levels of readout noise and demonstrated that the qGAN model can be successfully trained
even in the presence of this noise, although the common intrinsic instability of qGAN training cannot be overcome. We then examined the impact of different hyperparameters on the training and found that the learning rate of the quantum generator has the strongest impact, which becomes more pronounced with increasing readout-error rates. We also demonstrated that readout-error probabilities smaller than $p\sim 6\%$ do not significantly impact the qGAN training, while slightly larger readout errors still yield training convergence but substantially worse training results. To enhance these results, we studied the impact of two different readout-error mitigation methods, which both improved the qGAN training to a performance similar to the noise-free case. For two-qubit gate errors, we found that their effect on qGAN training is even more pronounced, preventing the convergence of training already for small error rates of $p\sim 1.5\%$. These results demonstrate that the mitigation of readout and gate errors plays a crucial role for qGAN training on current NISQ devices.

In future work, we will extend our current study to qGAN training on real quantum hardware.
Moreover, we will investigate mitigation techniques to tackle the common issue of training instabilities for qGAN models. Finally, we are planning to study the impact of quantum noise and error mitigation methods on other QML models beyond qGANs.

 \ack{This project is supported by CERN's Quantum Technology Initiative. L.F.\ is supported by the U.S.\ Department of Energy, Office of Science, National Quantum Information Science Research Centers, Co-design Center for Quantum Advantage (C$^2$QA) under contract number DE-SC0012704, by the DOE QuantiSED Consortium under subcontract number 675352, by the National Science Foundation under Cooperative Agreement PHY-2019786 (The NSF AI Institute for Artificial Intelligence and Fundamental Interactions, http://iaifi.org/), and by the U.S.\ Department of Energy, Office of Science, Office of Nuclear Physics under grant contract numbers DE-SC0011090 and DE-SC0021006. S.K.\ acknowledges financial support from the Cyprus Research and Innovation Foundation under project ``Future-proofing Scientific Applications for the Supercomputers of Tomorrow (FAST)'', contract no.\ COMPLEMENTARY/0916/0048.}

\section*{References}
\bibliographystyle{unsrt}
\bibliography{biblio.bib}
\end{document}